\documentstyle[11pt,epsfig]{article}
\begin{document}
\begin{center}
{\LARGE \bf Atmospheric Secondary Particles \\ In Near Earth Space \\}
~\\
Ming-Huey A. Huang \\
{\em Institute of Physics, Academia Sinica, Taipei, Taiwan, 11529, 
Republic of China  \\
E-mail: huangmh@phys.sinica.edu.tw }\\
~\\
Presented in \\
{\bf  The 8th Asia Pacific Physics Conference}
\footnote{Proceeding of the 8th Asia Pacific physics conference will be 
published by World Scientific publishing Co..}\\
Session BB-4 \\
7 Aug. 2000 \\
~\\
\begin{minipage}[t]{140mm}
The Alpha Magnetic Spectrometer detects a large amount of particles below 
rigidity cutoff. Those high energy particles create questions related to 
radiation belts and atmospheric neutrinos. To understand the origin of these 
particles, we use a trajectory tracing program to simulate particle 
trajectories in realistic geomagnetic field. The complex behaviors and large 
$e^+/e^-$ are explained here.
\end{minipage}
\end{center}
~\\
\section{Introduction}
In June 1998, the Alpha magnetic spectrometer AMS~\cite{AMS-NIM} had a test 
flight on board space shuttle and recorded some $10^8$ events. The AMS proton 
and leptons spectrum~\cite{ams-proton,ams-lepton} have a magnetic latitude 
dependence. The spectral shape of primary cosmic rays can be explained by the 
geomagnetic rigidity cutoff. However, below the rigidity cutoff, a second 
spectrum was clearly seen. The energy of this second spectrum is as high as 6 
GeV, 10 times higher than previous measured energy in the radiation belts! 
Beside this apparent difference, these second spectrum particles have similar 
trajectories as particles in trapped radiation, however, the fate are quite 
different. The discovery of these GeV sub-rigidity particles create new 
questions about the radiation belts and the atmospheric neutrinos flux. In 
this article, the trajectories of these particles are simulated in the same 
condition as AMS. They are shown to explain the complex behaviors of these 
secondary particles.

\section{Trajectory Tracing}
A trajectory tracing program~\cite{CJP-tracing} are used to trace particle 
trajectory forward and backward in time. The interactions between particles 
and atmosphere are ignored. The particles are defined as cosmic rays if their 
radial distance greater than $10R_E$ ($R_E$ mean Earth radius 6371.2km) when 
traced backward. 
The particle hit the ground in backward tracing is defined as atmospheric 
secondary. Physically, it is impossible to hit the ground at these energy 
regions. A limit of 40-km altitude was set as the top of atmosphere (TOA). The 
position where particle originates (traced backward) at 40 km altitude is 
defined as the {\em ``source''}. The position where particle re-enter 
atmosphere (traced forward) at 40 km altitude is defined as the 
{\em ``sink''}. The total time from the source to the sink, i.e. from creation 
to absorption by atmosphere, is called the {\em ``lifetime''}. Two groups of 
events are found, the Short-Lived Particles (SLP) which have lifetime in the 
order of bouncing period and the Long-Lived Particles (LLP) which have lifetime
 much longer than bounce period and less than drift period.

\section{The Short-Lived Particles}
\begin{figure}[htb]
\epsfxsize=35pc
\epsfbox{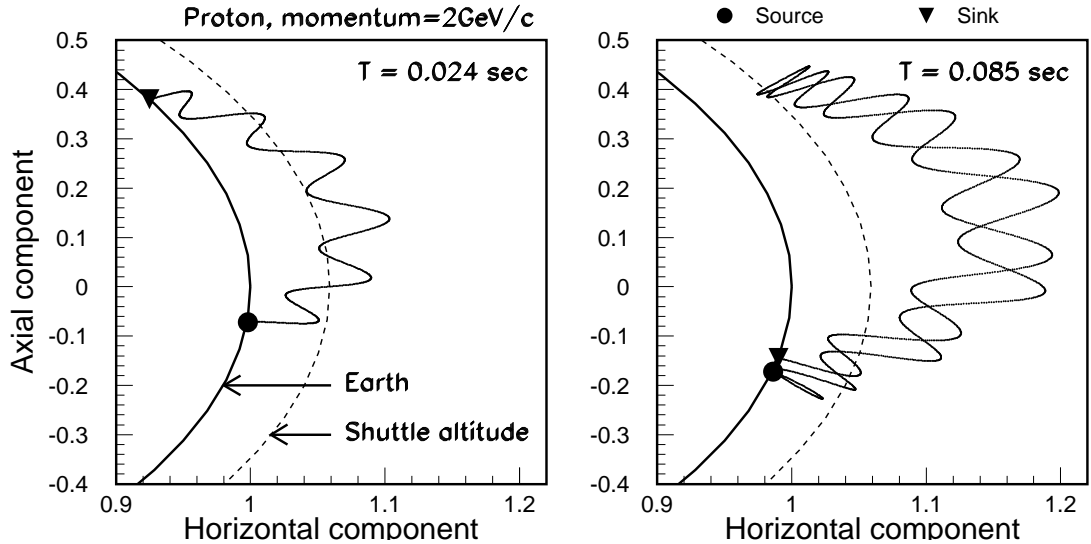}
\caption{Two examples of short-lived particle trajectories. The horizontal 
axis is the equatorial component and the vertical axis is the Earth rotational 
axial component. The left/right figures show an event which cross the magnetic 
equator once/twice. The lifetime from source (dot) to sink (triangle) is 
shown on the upper right corner.\label{fig:sft}}
\end{figure}

The SLP are absorbed by atmosphere in several bouncing motions between 
southern and northern hemisphere after their creation. Their motions can be 
categorized by the number of magnetic equator crossing times. This number 
corresponds to the ratio of lifetime to the half of bouncing period. The most 
common crossing times are 1 and 2. Figure~\ref{fig:sft} show the trajectories 
of those two cases. 
The bouncing period is inverse proportional to particle 
velocity~\cite{Martin}.  At GeV range, the electron and positron move at 
$\beta\sim 1$. Therefore, the lifetime is independent of energy.

Since SLP are created and absorbed by atmosphere in very short time, their 
source and sink positions must follow the magnetic field line to TOA and 
ground. Therefore the distribution of sources and sinks have a pattern similar 
to the flight path of detector. The SLP are detected all over the AMS covered 
area, latitude $-51.5^o$ to $+51.5^o$. In the equatorial region, there is a 
gap in the source/sink of SLP. Inside this gap, the local magnetic field line 
can not reach AMS altitude, particles within this gap could not fly up to AMS 
altitude and be detected as SLP.
\begin{figure}[htb]
\epsfxsize=35pc
\epsfbox{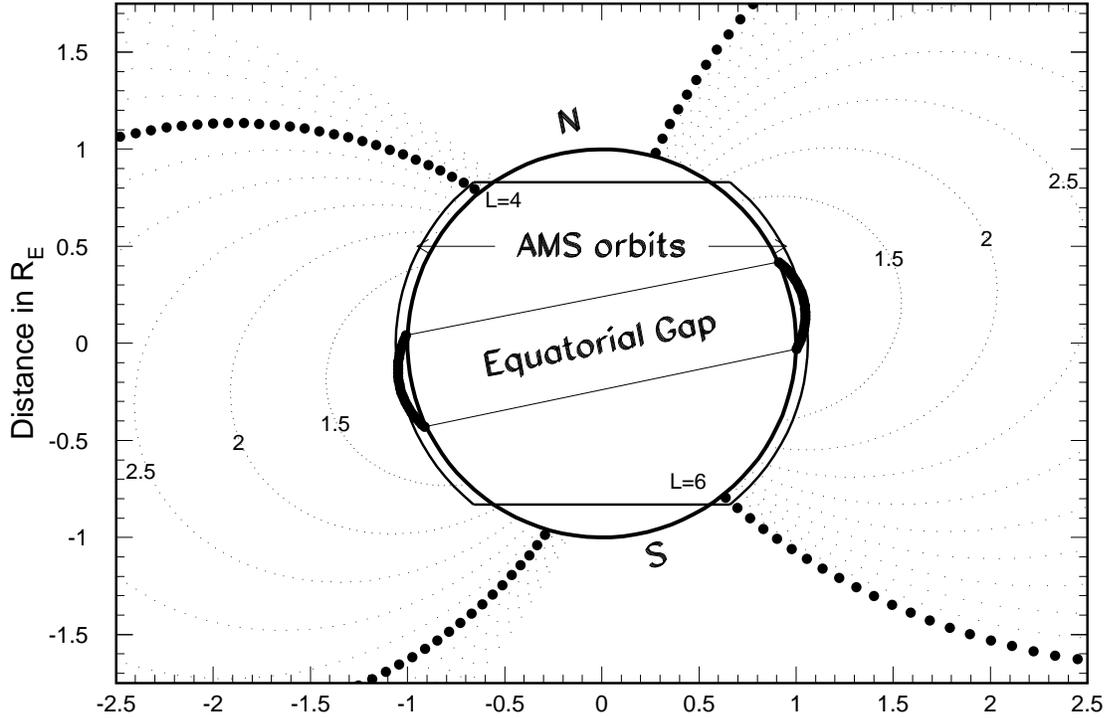}
\caption{The equatorial gap and space covered by short lived particles are 
shown in a slice along the dipole plane. Notice the short lived particles can 
cover space up to $4-6R_E$ in AMS covered region. Traditionally, radiation 
zone covers from $1.2R_E$ to $\sim 7R_E$.~\label{fig:lsh}}
\end{figure}

\section{The Long-Lived Particles}
\begin{figure}[hp]
\epsfxsize=35pc
\epsfbox{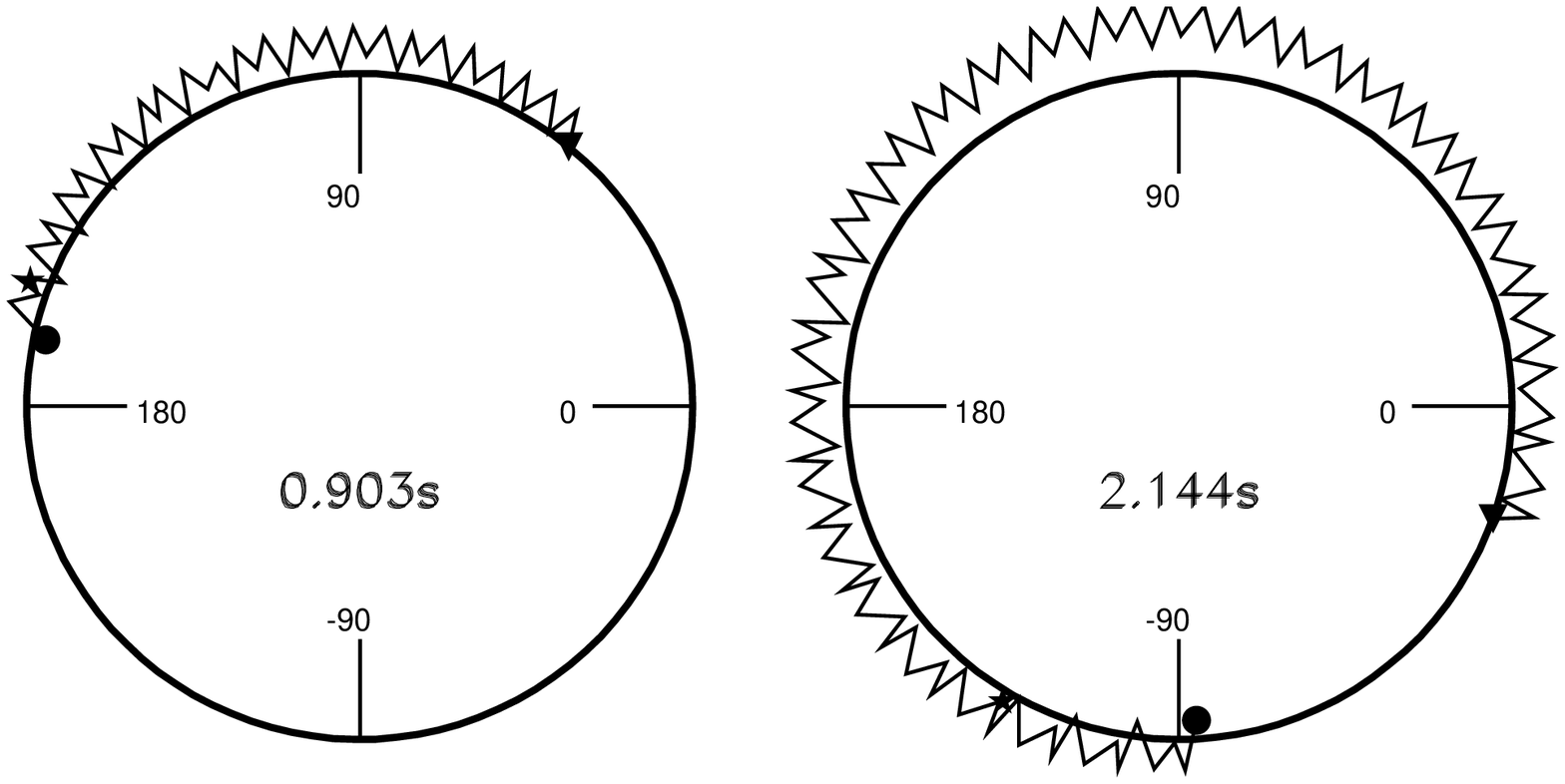}
\caption{Two examples of long-lived particles trajectories projected on the 
equatorial plane, viewing from North Pole. The circle is the Earth. Only the 
northern/southern mirroring points (those points closer to earth) and point of 
highest altitude (those points away from earth) are plotted. The total 
lifetime is also printed in the center of each figure.\label{fig:lft}}

\vspace{0.5cm}
\epsfxsize=35pc
\epsfbox{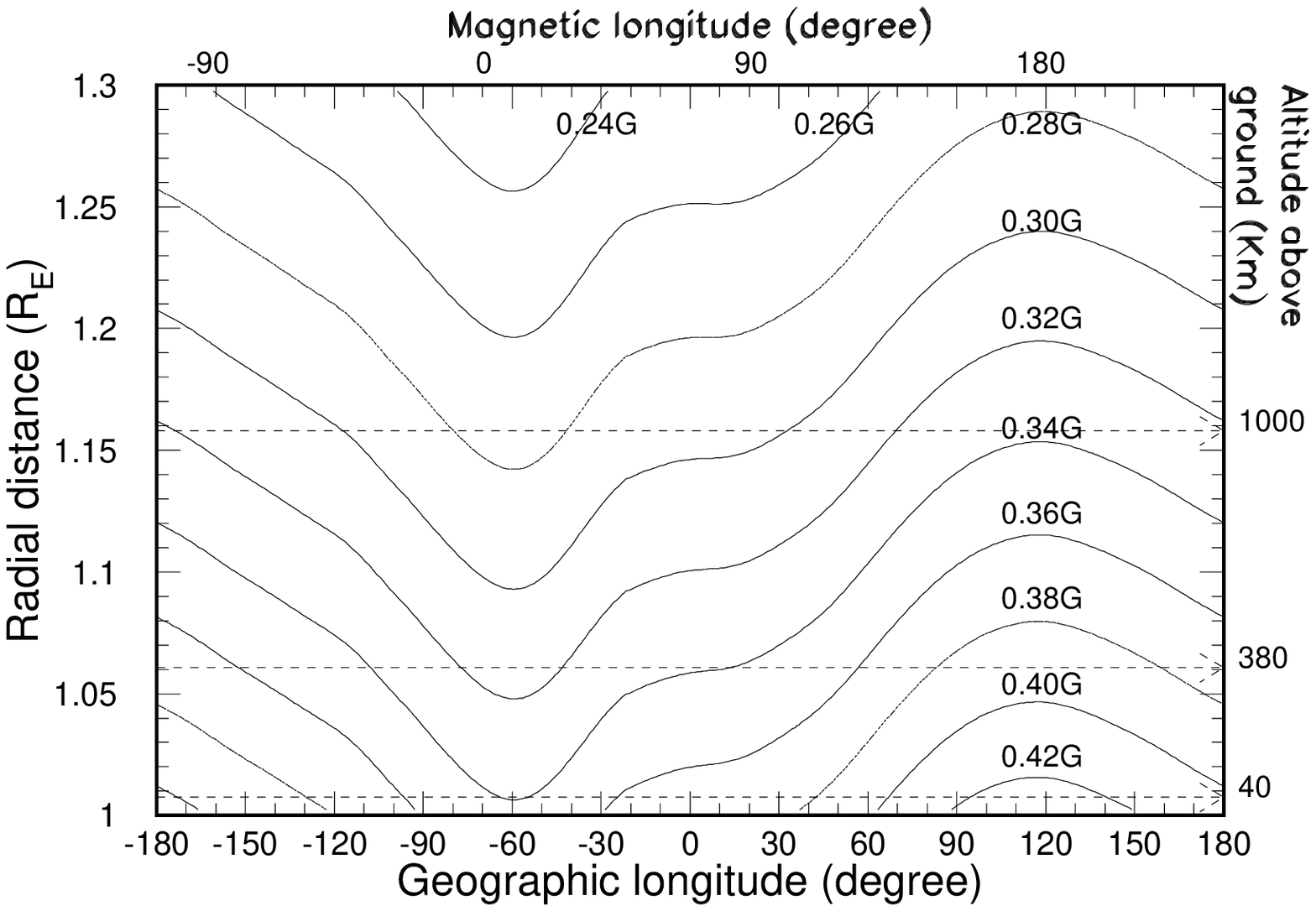}
\caption{The source and sink of long-lived particles are distributed at the 
east and west side of weakest magnetic field along the magnetic equator. The 
long-lived particles are created near magnetic longitude $0^o$, drift along 
the same L shell, pass the highest altitude near magnetic longitude $180^o$, 
then descend and enter atmosphere.\label{fig:bl}}
\end{figure}

Particles produced inside the gap need to drift to higher altitude to be 
detected. This drift motion increase the lifetime. Figure~\ref{fig:lft} show 
the summary of trajectories of two LLP events. The source and sink of the LLP are
 separated to two regions. The main reason is the offset of the center of 
geomagnetic dipole from the center of the Earth. Along the magnetic equator, 
one side (magnetic longitude $180^o$) the magnetic field are stronger, so the 
particle move to higher altitude; on the opposite side (magnetic longitude 
$0^o$), the magnetic field are weaker and particles move to lower altitude and 
could hit the TOA, shown in figure~\ref{fig:bl}. Positive LLP are generated in 
magnetic longitude $-180^0$ to $0^o$ and sink to magnetic longitude $0^o$ to 
$180^o$. Negative particles reverse this direction.
For particles coming directly 
from top of magnetic longitude $0^o$, they fly to AMS altitude not far from 
their source or sink. These regions are excluded in the AMS analysis. 
Therefore, the source and sink are separated in east and west of the magnetic 
longitude $0^o$.

The mean lifetime of LLP is approximately half of drift period which is 
proportional to $\sim 1/\gamma\beta^2$~\cite{Martin}. Therefore 
the lifetime and energy have an approximately inverse proportional relation. 

These LLP create a {\bf partial} ring current, their trajectories cover 
altitude from $1R_E$ to $1.2R_E$, the mirroring magnetic latitudes are within 
$\pm 25^o$. This {\bf non-uniform},(not {\bf anisotropic} as AMS claimed),
distribution of LLP is result of multi-pole moment of geomagnetic filed. In a 
centered dipole field, there will be no LLP. 

The source(sink) of positive(negative) LLP are separated to two groups, north 
and south of magnetic equator. This is the effect of asymmetry of geomagnetic 
field in magnetic latitude. The magnetic equator is not always the position 
where the magnetic field is minimum. At magnetic longitude $-60^o \sim 130^o$, 
the positions of minium magnetic field are south of magnetic equator, 
particles reach lower altitude in area south of magnetic equator and have 
chance to hit TOA. For magnetic longitude $>130^o$ or $<-60^o$, particles 
will hit TOA in area north of magnetic equator. 

\section{Position Electron Ratio}
\begin{figure}[htb]
\epsfxsize=35pc
\epsfbox{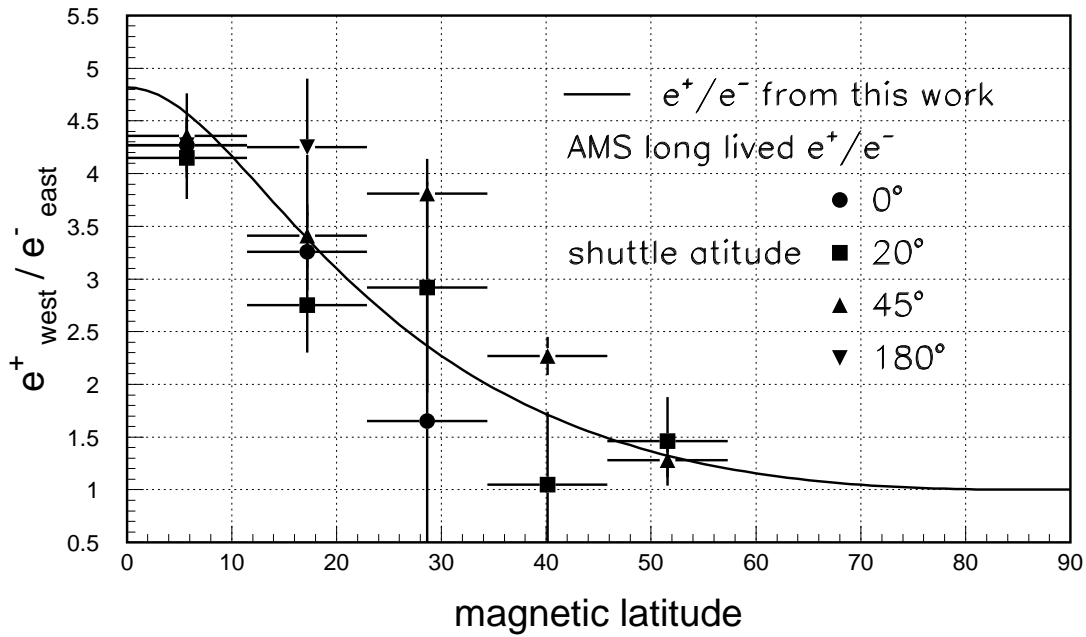}
\caption{The ratio of positron over electron from this work is consistent with 
the AMS long-lived measurements~\cite{ams-lepton}. This ratio drop to 1 at 
high latitude region where balloon experiment measure secondary 
$e^+/e^-\simeq 1$.\label{fig:ratio}}
\end{figure}

The secondary positron and electron ratio can be higher than 4 and have a 
latitude dependence. This effect is dominated by rigidity cutoff. Secondary 
positrons from west and electrons from east have chance to move to space. 
However, the rigidity cutoff from west is lower than from east, so 
there are more positrons than electrons. The difference of rigidity cutoff 
between east and west decrease at higher latitude, therefore the $e^+/e^-$ 
decreases too. Base on the following assumptions:
\begin{enumerate}
\item Positrons all come from west and electrons all come from east.
\item Dipole field and Stromer rigidity cutoff.
\item Primary cosmic rays flux follows power law spectrum 
	$\Phi(E)\propto E^{-\gamma}$. 
\item At interesting energy region $10GeV/c^2 < E < 1TeV/c^2$, the 
multiplicity of total secondary particles are simplified as proportional to 
power law of primary cosmic ray energy~\cite{multi},
	$M(E)\propto E^{\eta}$. 
\end{enumerate}
this simple model predict
\[ e^+/e^- = \frac{\Phi(e^+)}{\Phi(e^-)}
	=\left( \frac{1+\sqrt{1-\cos^3\lambda}}
		{1+\sqrt{1+\cos^3\lambda}} \right)^{2(-\gamma+\eta+1)}	\]
 The particles are produced in 40Km altitude and transported to 380Km by 
the same L shell. Using $\gamma=2.75$ and $\eta=0.5$, the predicted  $e^+/e^-$,
 shown in figure~\ref{fig:ratio}, are consistent with the AMS results. 

For different incident angles, the difference in 
rigidity cutoff decrease, so does $e^+/e^-$. However, the conservation of 
isospin make the positron multiplicity higher than electron multiplicity at 
primary energy below $10 GeV/c^2$. This effect will increase $e^+/e^-$.  
A Monte-Carlo simulation which include realistic geomagnetic field and 
particle interactions is in progress.

\section*{Acknowledgments}
This work is supported by NSC-89-2112-M-001-034, National Science Council, Taiwan, Republic of China.


\begin{thebibliography}{99}
\bibitem{AMS-NIM} S. Ahlen. {\it et al}, {\em Nucl. Instrum. Methods}  
	{\bf A 350}, 351, (1994). 
\bibitem{ams-proton} AMS collaboration, J. Alcaraz,  {\it et al}, 
	{\em Phys. Lett.}  {\bf B 472}, 215, (2000)
\bibitem{ams-lepton} AMS collaboration, J. Alcaraz, {\it et al}, 
	{\em Phys. Lett.}  {\bf B 484}, 10, (2000)
\bibitem{CJP-tracing} M.A. Huang, {\it et al}, submitted to 
	{\em Chinese J. of Phys.} {\bf 38}, ,(2000)
\bibitem{Martin}Martin Walt {\em Introduction to Geomagnetically Trapped 
Radiation}, (Cambridge Univ. Press, 1994).
\bibitem{multi} 
	J. Benecke, {\it et al}, {\em Nucl. Phys.} {\bf B 76}, 29, (1976); 
	W.M. Morse, {\it et al}, {\em Phys. Rev.} {\bf D 15}, 66, (1977);
	A. Breakstone, {\it et al}, {\em Phys. Rev.} {\bf D 30}, 528, (1984);
\end{thebibliography}
\end{document}